\documentclass[aps,prb,twocolumn,groupedaddress,showpacs,floatfix]{revtex4}
\usepackage{graphicx}
\begin{document}
\pagenumbering{arabic}
\pagestyle{plain}
\title{Investigation of the magnetic dipole field at the atomic scale in quasi-one-dimensional paramagnetic conductor Li$_{0.9}$Mo$_{6}$O$_{17}$}  
\author{Guoqing Wu,$^{1,3}$ Bing Wu,$^{2}$ and W. G. Clark$^{3}$}
\affiliation{$^{1}$College of Physics Science and Technology, Yangzhou University, Yangzhou, Jiangsu 225002, China}
\affiliation{$^{2}$Department of Math and Computer Science, Fayetteville State University, Fayetteville, NC 28301, USA}
\affiliation{$^{3}$Department of Physics and Astronomy, University of California, Los Angeles, California 90095, USA}
\altaffiliation{Electronic address: gwu999@gmail.com (Guoqing Wu)}
\date{\today}
\begin{abstract}
    We report magnetic dipole field investigation at the atomic scale in a single crystal of quasi-one-dimensional (Q1D) paramagnetic conductor Li$_{0.9}$Mo$_{6}$O$_{17}$, using a paramagnetic electron model and $^{7}$Li-NMR spectroscopy measurements with an externally applied magnetic field $B_{0}$ = 9 T. We find that the magnetic dipole field component ($B_{||}^{\text{dip}}$) parallel to $B_{0}$ at the Li site from the Mo electrons has no lattice axial symmetry; it is small around the middle between the lattice $c$ and $a$ axes in the $ac$-plane with the minimum at the field orientation angle $\theta$ = +52.5$^{\circ}$, while the $B_{||}^{\text{dip}}$ maximum is at $\theta$ = +142.5$^{\circ}$ when $B_{0}$ is applied perpendicular to $b$ ($B_{0}$ $\perp$ $b$), where $\theta$ = 0$^{\circ}$ represents the direction of $B_{0}$ $\parallel$ $c$. Further estimate indicates that $B_{||}^{\text{dip}}$ has a maximum value of 0.35 G at $B_{0}$ = 9 T, and the Mo ions have a possible effective magnetic dipole moment 0.015 $\mu_{\text{B}}$ per ion, which is significantly smaller than that of a spin 1/2 free electron. By minimizing potential magnetic contributions to the NMR spectra satellites with the NMR spectroscopy measurements at the direction where the value of the magnetic dipole field is the smallest, the behavior of the independent charge contributions is observed. This work demonstrates that the magnetic dipole field from the Mo electrons is the dominant source of the local magnetic fields at the Li site, and it suggests that the mysterious ``metal-insulator'' crossover at low temperatures is not a charge effect. The work also reveals valuable local field information for further NMR investigation which is suggested recently [Phys. Rev. B $\bf{85}$, 235128 (2012)] to be key important to the understanding of many mysterious properties of this Q1D material of particular interest.   
%
\end{abstract}
\maketitle
\section{Introduction}
    The quasi-one-dimensional (Q1D) paramagnetic conductor, Li$_{0.9}$Mo$_{6}$O$_{17}$, has been of particular interest because of its unusual properties. It is thought to exhibit transport properties associated with a Luttinger liquid \cite{chud, giam, hager} at high temperatures, and otherwise many of its properties have long been mysterious \cite{green, mcc,sepper,nuss,cohn}. Among these is an unusual increase (an upturn) in resistivity \cite{green,dumas,jfm}, which shows a ``metal-insulator'' crossover at low temperatures (the crossover temperature $T_{\text{MI}}$ = 24 K), for which a robust explanation remains elusive, while four completely different mechanisms were theoretically proposed\cite{green1,green2}: charge-density wave (CDW), spin-density wave (SDW), localization, and Luttinger liquid. It is also a superconductor (``insulator''-superconductor transition temperature $T_{c}$ = 2.2 K), which is most recently\cite{jfm, sepper} found to be three-dimensional (3D). Thus it involves an electron dimensional crossover,\cite{john1, john4} and may also involve spin triplet Cooper pairs \cite{sepper, lebed}, with a triplet superconducting state as one of its potential electron ground states.  

    Li$_{0.9}$Mo$_{6}$O$_{17}$ has also been a subject of intensive experimental studies over the last three decades \cite{green, mcc}. Many different types of experiments has been performed: x-ray diffraction \cite{onoda}, neutron scattering \cite{john1}, dc magnetic susceptibility \cite{choi,jfm1, matsuda}, resistivity \cite{jfm, choi, xxu, msd}, heat capacity \cite{schlen}, thermal expansion \cite{john4}, thermal conductivity \cite{wake,cohn}, Nernst signal \cite{cohn1}, optical conductivity \cite{choi}, muons spectroscopy \cite{chak}, scanning tunneling microscope (STM) \cite{hager}, and angle resolved photo-emission (ARPES) \cite{denl}. But it faces tremendous challenges over these experiments for the physics understanding. For example, for the ``metal-insulator'' crossover at 24 K, x-ray diffraction and neutron scattering show no evidence of structure phase transition \cite{onoda, john1}, dc susceptibility shows no signs of Curie-Weiss deviation in the electron magnetization \cite{choi, jfm1, matsuda}, and heat capacity indicates negligible associated heat anomaly \cite{schlen}.   

    Because of these challenges and the limitations in many of these experimental techniques, other capable experimental techniques are highly valuable. For example, most recent theoretical studies in Li$_{0.9}$Mo$_{6}$O$_{17}$ have strongly suggested \cite{merino, nuss} the significance of charge and spin fluctuations $\&$ correlation which are related to the local electric and magnetic fields arising from the Mo electrons and could be measured at the Mo or nearby atom site like the Li. However, none of above experimental techniques were able to probe them directly at the atomic scale. On the other hand, in terms of a Luttinger liquid (if this is the case), long-range Coulomb interactions among the conduction electrons are expected \cite{chud, giam, nuss} to be stronger than those in a traditional Fermi liquid. The interaction could induce electron polarizations and thus have a direct impact on the local electric and magnetic fields. Therefore, it is important to investigate the local electric and magnetic field from the Mo electrons. Moreover, the field reveals the sources of the charge and spin dynamics \cite{slichter, abrag} of the Mo electrons at the atomic scale.  

    Nuclear magnetic resonance (NMR) is a well-known versatile local probe capable of directly measuring the local electric and magnetic field including the electron charge and spin dynamics at the atomic scale \cite{slichter, abrag}. 

    In this paper, we report local electric and magnetic field investigation, using a theoretical paramagnetic electron model and $^{7}$Li-NMR spectroscopy measurements on a single crystal of Li$_{0.9}$Mo$_{6}$O$_{17}$, with an externally applied magnetic field $B_{0}$ = 9 T. Since we expect the magnetic dipole field from the paramagnetic Mo electrons to be one of the major local field sources (at least one of them) at the Li site according to the NMR theory \cite{slichter, abrag}, the magnetic dipole field is the focus in this investigation.     

    In fact, magnetic dipole field that originated from the magnetic dipole moments of the electron spins, including the unpaired spins of the paramagnetic conduction electrons, are of particular interest in various aspects of NMR, including NMR spectroscopy, Knight shift, spin-lattice relaxation, and spin-echo decay, especially when the dipolar hyperfine couplings to the electron spins are significant, or when the time scale of their fluctuations matches that of the dynamics for the spin-lattice relaxation or spin-echo decay rates in the materials \cite{abrag, slichter, carter}. For example, a NMR spectrum could be inhomogeneously broadened and a Knight shift could have a significant value, due to the contribution of the magnetic dipole fields from the electron spins \cite{abrag, slichter, carter, wu2}. Similar effect could also be generated by other local field sources at a nucleus when they are not negligible \cite{dick1, drain, dick2}. Unlike other local field sources, magnetic dipole field is always associated with the size $\&$ orientation of the magnetic dipole moments, and depends on the dipolar hyperfine coupling between the nucleus and the electron spins at the atomic scale, which can be estimated from the structure of the crystal lattice theoretically and can also be measured by the NMR techniques experimentally. 



    Our main results are that the magnetic dipole field component ($B_{||}^{\text{dip}}$) parallel to $B_{0}$ from the Mo electrons at the Li site is found to have no lattice axial symmetry; it is small around the middle between the lattice $c$ and $a$ axes in the $ac$-plane, $\sim$ 7.5$^{\circ}$ closer to the $a$-axis for the central minimum which is at $\theta$ = +52.5$^{\circ}$, and the maximum is at $\theta$ = +142.5$^{\circ}$ when $B_{0}$ is applied perpendicular to $b$ ($B_{0}$ $\perp$ $b$) (note, $\theta$ is one of the orientation angles of $B_{0}$, and $\theta$ = 0$^{\circ}$ represents the direction of $B_{0}$ $\parallel$ $c$). Our further estimate indicates that the maximum value of $B_{||}^{\text{dip}}$ at the Li site is $\sim$ 0.35 G when $B_{0}$ = 9 T $\perp$ $b$, and the Mo ions have a possible effective magnetic dipole moment ($\mu_{\text{eff}}$) of 0.015 $\mu_{\text{B}}$, which is significantly smaller than that of a spin $S$ = 1/2 free electron. By separating the charge contribution from a mixture of major magnetic contributions to the NMR spectra satellites with the NMR spectroscopy measurements at the direction where the value of the magnetic dipole field is the smallest (i.e., the magnetic contribution is minimized), the behavior of the independent charge contributions is observed.

    This work demonstrates that the magnetic dipole field from the Mo electrons is the dominant source of the local magnetic fields at the Li site, and suggests that the mysterious ``metal-insulator'' crossover at low temperatures is not a charge effect. The work also reveals valuable local electric and magnetic field information for further NMR investigation which is strongly suggested \cite{merino} recently to be key important to the understanding of many mysterious properties \cite{green, choi, lebed} of this Q1D material.
 
    The rest of the paper is organized as follows. First, Section II presents the calculation of the magnetic dipole field at the atomic scale in paramagnetic electron systems, which is described in a general form so that the method can be used for applications in other electron systems, and the field in the Q1D paramagnetic conductor Li$_{0.9}$Mo$_{6}$O$_{17}$ is calculated when a single crystal sample is exposed to an externally applied magnetic field $B_{0}$. Second, Section III has the experimental result of the $^{7}$Li-NMR spectra corresponding to the result of the theoretical calculations obtained in Section II. Third, Section IV presents discussions regarding the electron model used in the study and related physics quantities. Finally, the conclusions are stated in Section V.     
\section{Calculation of the magnetic dipole field at the atomic scale}
    As illustrated in Fig. 1 (a), a magnetic dipole moment $\vec{\mu}_{j}$ from the electrons of an atom at the site M (electron moment site $j$) in the crystal lattice can produce a magnetic dipole field ($\vec{B}_{ij}$) at a nearby atom site P (field observation site $i$). The value of $\vec{B}_{ij}$ is given by \cite{jackson}
\begin{equation}
\vec{B}_{ij} = \frac{\mu_{0}}{4 \pi}\left[\frac{3~\vec{r}_{ij}(\vec{\mu}_{j}\cdot\vec{r}_{ij})}{r_{ij}^{~5}} - \frac{\vec{\mu}_{j}}{r_{ij}^{~3}}\right],
\end{equation}
where $\vec{r}_{ij}$ is the displacement vector from $j$ to $i$ (M $\rightarrow$ P), and $\mu_{0}$ is the permeability constant. 
\begin{figure}
\includegraphics[scale= 0.34]{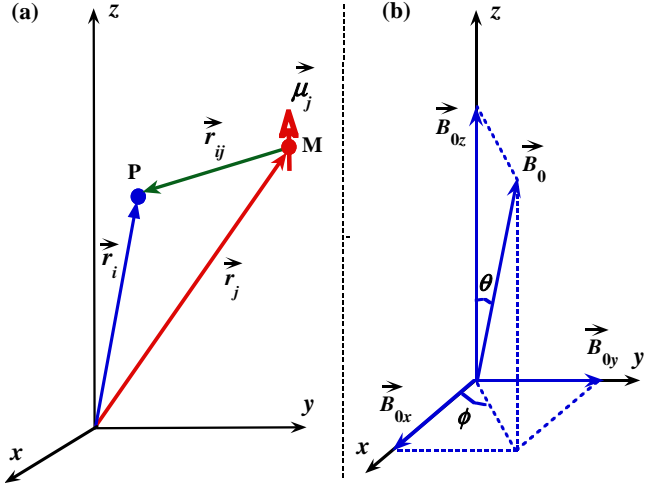}
\caption{(color online) (a) Cartesian coordinate system, where an electron magnetic dipole moment $\vec{\mu}_{j}$ is at the atom site M near the field observation site P in the crystal lattice. (b) The orientation of the externally applied magnetic field, $\vec{B_{0}}$. \label{fig1}}  
\end{figure}  

    The total dipolar field ($\vec{B}$) at the observation site P is the summation of the field from the moments at all the moment sites $j$,
\begin{eqnarray}
  \vec{B} & \equiv & <\vec{B}_{i}> ~= ~\sum_{j}<\vec{B}_{ij}>, \\
                & = & \frac{\mu_{0}}{4 \pi} \sum_{j}<\left[\frac{3~\vec{r}_{ij}(\vec{\mu}_{j}\cdot\vec{r}_{ij})}{r_{ij}^{~5}} - \frac{\vec{\mu}_{j}}{r_{ij}^{~3}} \right]>, \\
                & = & \frac{\mu_{0}}{4 \pi} \sum_{j}\left[\frac{3~\vec{r}_{ij}(<\vec{\mu}_{j}>\cdot\vec{r}_{ij})}{r_{ij}^{~5}} - \frac{<\vec{\mu}_{j}>}{r_{ij}^{~3}} \right]. 
\end{eqnarray}
The magnetization $\vec{M}$ due to the electron moments is 
\begin{equation}
   \vec{M} = \sum_{j}<\vec{\mu}_{j}> / V,
\end{equation} 
where $V$ is the sample volume.

    For paramagnetic electrons, $\vec{M}$ essentially has a very accurate linear dependence\cite{ash} on the externally applied magnetic field $\vec{B}_{0}$, i.e.,
\begin{equation}
   \vec{M} = \chi \vec{H}_{0},  ~~\text{and} ~~ \vec{H}_{0} = \vec{B}_{0}/\mu_{0},
\end{equation}
where $\vec{H_{0}}$ is the intensity of the applied magnetic field, and $\chi$ is the sample paramagnetic susceptibility (isotropic), which can be a $T-$dependent variable [the deviation from their linear relation in Eq. (6) is in the order of $\sim$ $\chi^{2}\vec{H}_0$, which is negligible as $\chi$ $<$ $\sim$ 10$^{-3}$ (cm$^{3}$/mol.ion) for most known materials]. If magnetic anisotropy is considered, then $M_{i'}$ = $\chi_{i'j'} H_{0j'}$, where $\chi_{i'j'}$ is the tensor element of the susceptibility ($\chi_{i'j'}$ = $\partial M_{i'}$/$\partial H_{0j'}$) and $i', ~j'$ = $x, ~y, ~z$.

    Thus considering Eqs. (1)-(6), we have
\begin{equation}
  \vec{B} = \frac{\mu_{0}}{4 \pi} ~\chi V \sum_{j=-N}^{+N} \left[\frac{3\vec{r}_{ij}(\vec{B}_{0}\cdot\vec{r}_{ij})}{r_{ij}^{~5}} - \frac{\vec{B}_{0}}{r_{ij}^{~3}} \right], \\
\end{equation}
where $N$ in the index $j$ is the number of sites for the electron moments. Equation (7) gives the $x$, $y$ and $z$ dipolar field components at the observation site $i$ as
\begin{widetext}
\begin{eqnarray}
  B_{x} = \frac{\mu_{0}}{4 \pi} ~\chi V \sum_{j=-N}^{+N} \left[\frac{3{x}_{ij}(B_{0x}x_{ij} + B_{0y}y_{ij} + B_{0z}z_{ij})}{r_{ij}^{~5}} - \frac{B_{0x}}{r_{ij}^{~3}} \right], \\
  B_{y} = \frac{\mu_{0}}{4 \pi} ~\chi V \sum_{j=-N}^{+N} \left[\frac{3{y}_{ij}(B_{0x}x_{ij} + B_{0y}y_{ij} + B_{0z}z_{ij})}{r_{ij}^{~5}} - \frac{B_{0y}}{r_{ij}^{~3}} \right], \\
  B_{z} = \frac{\mu_{0}}{4 \pi} ~\chi V \sum_{j=-N}^{+N} \left[\frac{3{z}_{ij}(B_{0x}x_{ij} + B_{0y}y_{ij} + B_{0z}z_{ij})}{r_{ij}^{~5}} - \frac{B_{0z}}{r_{ij}^{~3}} \right]. 
\end{eqnarray}
\end{widetext}

    Figure 1 (b) shows the orientation of the applied magnetic field $\vec{B}_{0}$, which can be expressed as $\vec{B}_{0}$ = $B_{0x}\hat{i}$ + $B_{0y}\hat{j}$ + $B_{0z}\hat{k}$ = $B_{0}(\sin\theta \cos\phi ~\hat{i} + \sin\theta \sin\phi ~\hat{j} + \cos\theta ~\hat{k}$), where $\theta$ and $\phi$ are the standard spherical angles in the Cartesian system.\cite{arfken} 

    Thus the dipolar field components $B_{x}$, $B_{y}$, and $B_{z}$ along the $x$, $y$, and $z$ directions, respectively, can be calculated with Eqs. (8)-(10) by considering the coordinates of all the electron moment sites (atom sites M) that are included (as many as possible).  

    When the values of $\vec{B}$ obeys $|\vec{B}|$ $<<$ $B_{0}$, the contribution of $\vec{B}$ to an NMR spectrum and Knight shift comes only from the component of $\vec{B}$ $\parallel$ $\vec{B_{0}}$ ($B_{||}^{\rm{dip}}$), which is also the case in our experimental observations with an applied magnetic $B_{0}$ = 9 T.     
\begin{figure}
\includegraphics[scale= 0.36]{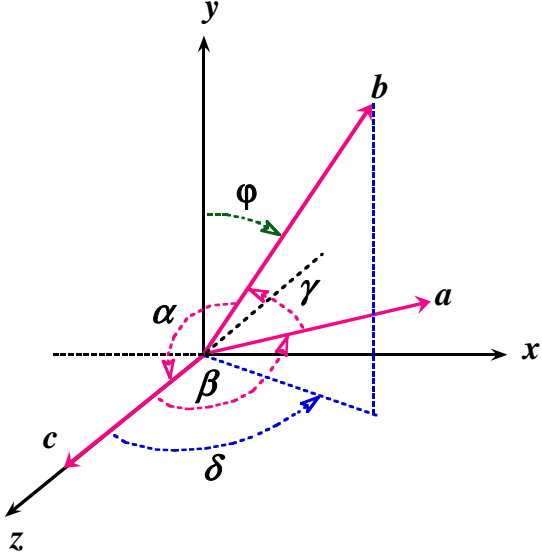}
\caption{(color online) A fixed set-up of the Cartesian $x$, $y$, and $z$ axes relative to the $a$, $b$, and $c$ axes of the crystal lattice in the lattice coordinate system that can always be made, where the $xz$-plane is in the $ac$-plane and the $z$-axis is along the $c$-axis. \label{fig2}}  
\end{figure}  

    The calculation using Eqs. (8)-(10) requires Cartesian coordinates, thus for coordinates given in the lattice $abc$-system a transform matrix ($M_{tran}$) is needed. This can be done by a fixed set-up of the Cartesian $x$, $y$, and $z$ axes versus the $a$, $b$, and $c$ axes of the crystal lattice in the lattice coordinate system that can always be made, where $xz$-plane is chosen to be placed in the $ac$-plane and the $z$-axis is along the $c$-axis, as illustrated in Fig. 2, from which we have 
\begin{eqnarray}
   \vec{e}_{a} & = & (sin\beta, ~0, ~cos\beta), \\
   \vec{e}_{b} & = & (sin\varphi sin\delta, ~cos\varphi, ~sin\varphi cos\delta), \\
   \vec{e}_{c} & = & (0, ~0, ~1), 
\end{eqnarray}
where $\vec{e}_{a}$, $\vec{e}_{b}$, and $\vec{e}_{c}$ are the Cartesian expression for the unit vectors of the $a$, $b$, and $c$ axes of the crystal lattice in the lattice coordinate system, respectively, and 
\begin{equation}
   \vec{e}_{a} \cdot  \vec{e}_{b} = cos\gamma, ~~\text{and} ~~  \vec{e}_{b} \cdot \vec{e}_{c} = cos\alpha. \\
\end{equation}   

   This gives
\begin{eqnarray} 
   sin\varphi sin\delta & = &  \frac{cos\gamma - cos\alpha cos\beta}{sin\beta}, \\
   sin\varphi cos\delta & = &  cos\alpha,  \\
   cos\varphi ~~~~~ & = &\sqrt{sin^2\alpha - \Big(\frac{cos\gamma - cos\alpha cos\beta}{sin\beta}\Big)^{2}}.  
\end{eqnarray}

   By considering Eqs. (11) - (17), we have \\
\begin{widetext}
\begin{equation}
\left(
\begin{array} {c}
     \vec{e}_{a} \\
     \vec{e}_{b} \\
     \vec{e}_{c} 
\end{array} 
\right) 
= 
\left(
\begin{array} {c}
     ~~~~sin\beta   ~~~~~~~~~~~~~~~~~~~~~~~~~~  0  ~~~~~~~~~~~~~~~~~~~~~~ cos\beta \\
     \frac{cos\gamma - cos\alpha cos\beta}{sin\beta} ~~~ \sqrt{sin^2\alpha - \Big(\frac{cos\gamma - cos\alpha cos\beta}{sin\beta}\Big)^{2}} ~~~ cos\alpha \\
     ~~~~0  ~~~~~~~~~~~~~~~~~~~~~~~~~~~~ 0  ~~~~~~~~~~~~~~~~~~~~~~~~~ 1 
\end{array} 
\right) 
\left(
\begin{array} {c}
     \vec{e}_{x} \\
     \vec{e}_{y} \\
     \vec{e}_{z}
\end{array} 
\right),
\end{equation}
or
\begin{equation}
\left(
\begin{array} {c}
     x, y, z \\
\end{array}
\right)
\left(
\begin{array} {c}
     \vec{e}_{x} \\
     \vec{e}_{y} \\
     \vec{e}_{z} 
\end{array} 
\right) 
= 
\left(
\begin{array} {c}
     sin\beta  ~~~~~~~~~~~~~~~~~ \frac{cos\gamma - cos\alpha cos\beta}{sin\beta} ~~~~~~~~~~~~~~~~~~ 0 \\
      ~~  0 ~~~~~~~~~~~ \sqrt{sin^2\alpha - \Big(\frac{cos\gamma - cos\alpha cos\beta}{sin\beta}\Big)^{2}}~~~~~~~ 0 \\
     cos\beta  ~~~~~~~~~~~~~~~~~~~~~~ cos\alpha  ~~~~~~~~~~~~~~~~~~~~~~~~ 1 
\end{array} 
\right) 
\left(
\begin{array} {c}
     x'_{a}, y'_{b}, z'_{c} \\
\end{array}
\right)
\left(
\begin{array} {c}
     \vec{e}_{a} \\
     \vec{e}_{b} \\
     \vec{e}_{c}
\end{array} 
\right),
\end{equation}
\end{widetext}
where $\alpha$, $\beta$, and $\gamma$ are the lattice constants (including the values of $a$, $b$ and $c$), $x'_{a}$, $y'_{b}$, and $z'_{c}$ are the atom coordinates in the lattice $abc$-coordinate system, and $\vec{e}_{x}$, $\vec{e}_{y}$, and $\vec{e}_{z}$ are the unit vectors of the Cartesian $x$, $y$, and $z$ axes, respectively. Thus with Eq. (18) or (19) the coordinates between the lattice and Cartesian coordinates systems are easily transformable.
\begin{figure}
\includegraphics[scale= 0.44]{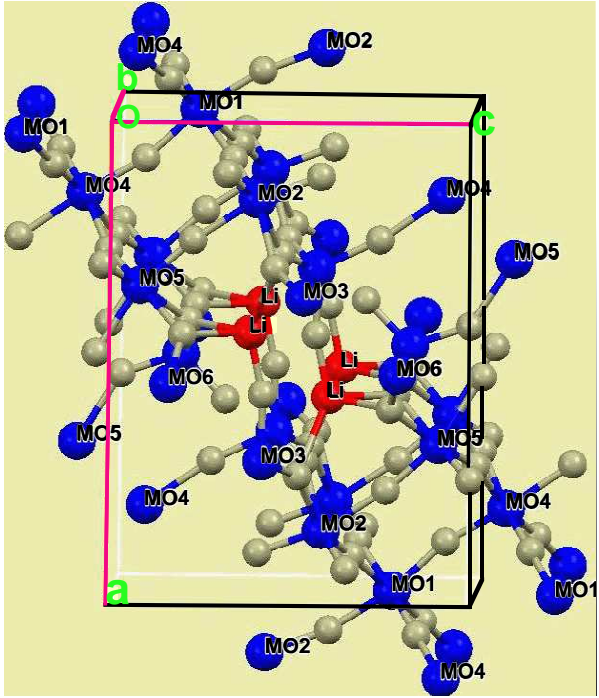} 
\caption{(color online) The crystal structure of Li$_{0.9}$Mo$_{6}$O$_{17}$ obtained from the neutron scattering data.\cite{john1, john2} All the Mo (blue color) and Li (red color) atoms in a unit cell are labeled, while the unlabeled ones (grey color) are the O (oxygen) atoms. The lattice $a$, $b$ and $c$ axes are specified with the lines in pink color. \label{fig3}}  
\end{figure}  
\begin{table*}
\caption{The fractional coordinates for the positions of the independent Mo and Li sites in the unit cell with the lattice constants in the lattice coordinate system (value at 300 K). [Ref. 16] \label{tab1}}
\begin{ruledtabular}
\begin{tabular}{ccccccc}
& atom sites & $x'$ & $y'$ & $z'$ \\ \hline
& Mo(1) & -0.00613 & 0.25 & 0.23356 \\ 
& Mo(2) & 0.14436 & 0.75 & 0.41840 \\ 
& Mo(3) & 0.31105 & 0.25 & 0.56755 \\ 
& Mo(4) & 0.16635 & 0.25 & -0.07938 \\ 
& Mo(5) & 0.31980 & 0.75 & 0.09404 \\ 
& Mo(6) & 0.49299 & 0.25 & 0.19604 \\
& Li & 0.40240 & 0.75 & 0.40904 \\ \hline
& $a$ = 12.762 $(\rm\AA)$ & & & \\
& $b$ = 5.523 $(\rm\AA)$  & & & \\
& $c$ = 9.499 $(\rm\AA)$  & & space group $P2_{1}/m$ (monoclinic) & \\
& $\alpha$ = 90$^{\circ}$ & & & \\
& $\beta$ = 90.61$^{\circ}$  & & unit cell volume $v$ = 669.5 $(\rm\AA)^{3}$ & \\
& $\gamma$ = 90$^{\circ}$ & & & \\
\end{tabular}  
\end{ruledtabular}
\end{table*}
%
%
\begin{figure}
\includegraphics[scale= 0.57]{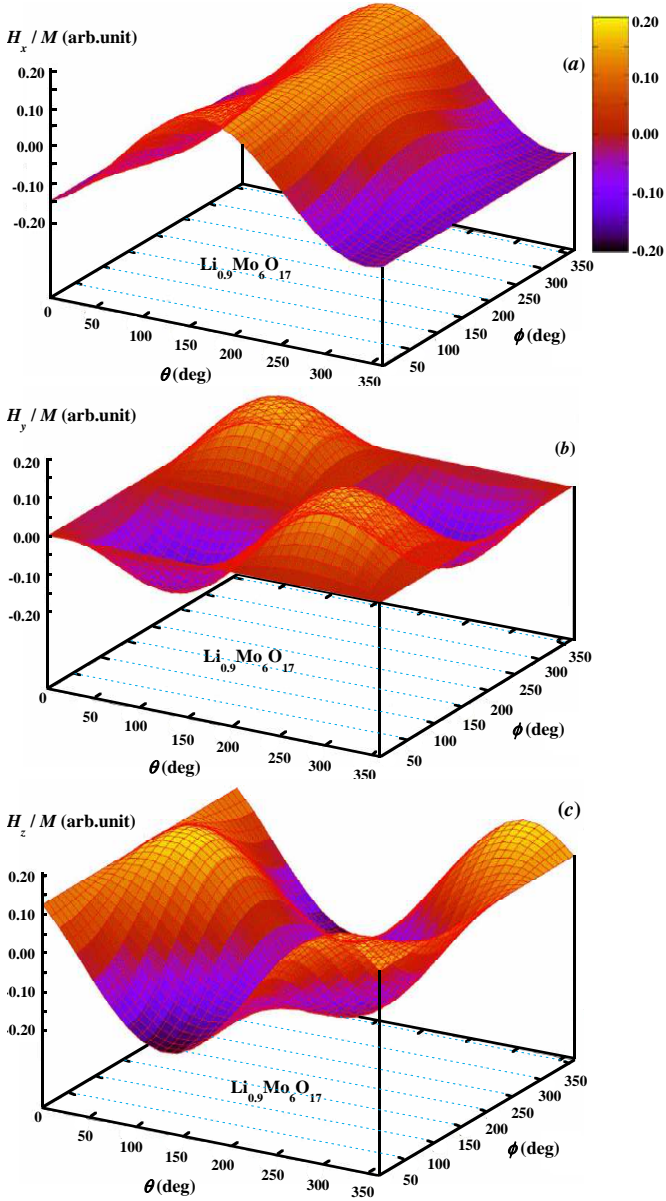}
\caption{(color online) Calculated magnetic dipole field components at the Li site in Li$_{0.9}$Mo$_{6}$O$_{17}$, as a function of the $\vec{B}_{0}$ orientation angles $\theta$ and $\phi$ in space, plotted as: (a) $H_{x}/M$, (b) $H_{y}/M$, and (c) $H_{z}/M$ versus $\theta$ and $\phi$ along the $x$, $y$ and $z$ axes, respectively. Here $M$ is the magnitude of the magnetization due to the electron paramagnetic moment. \label{fig4}}  
\end{figure}  
%
%
\begin{figure}
\includegraphics[scale= 0.64]{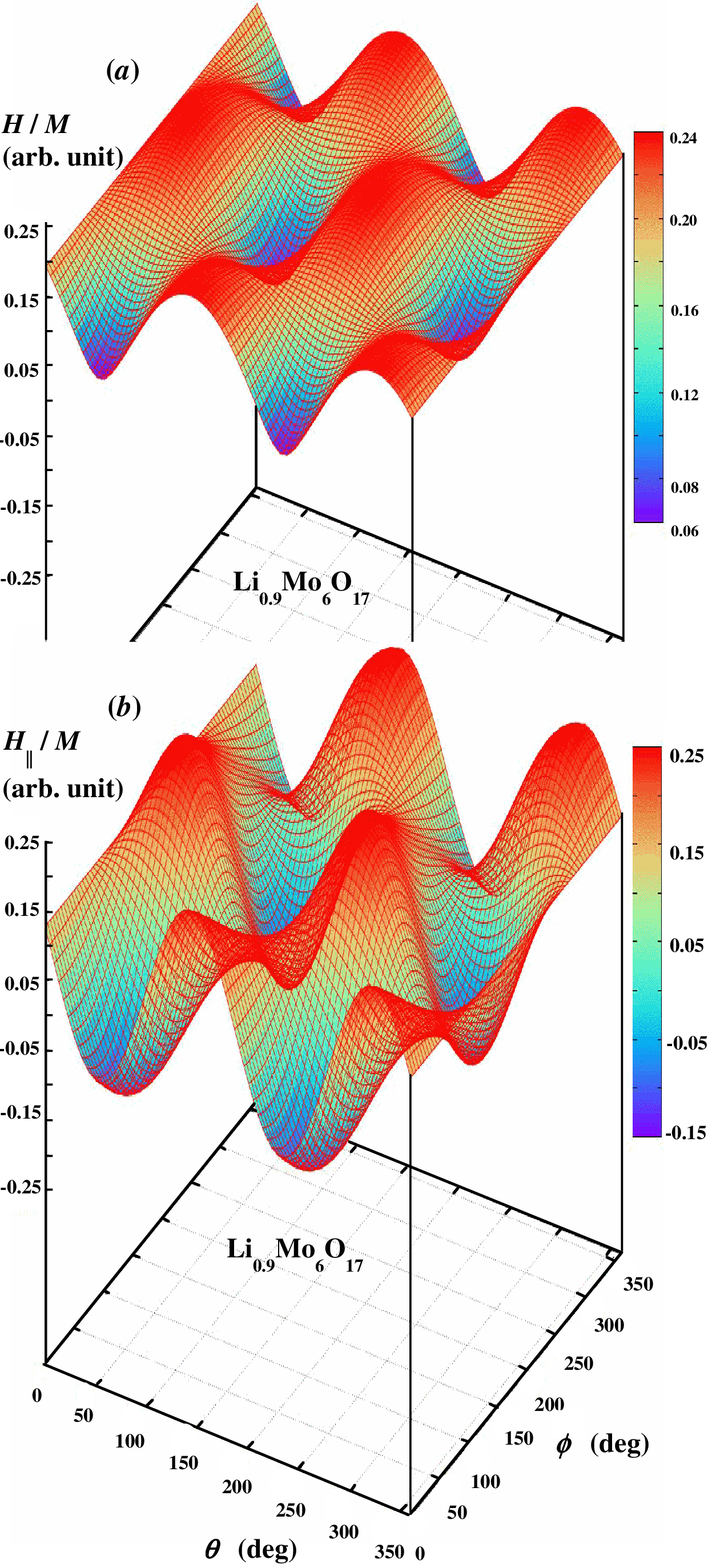}
\caption{(color online) Calculated (a) magnetic dipole field magnitude and (b) magnetic dipole field component parallel to $\vec{B_{0}}$ at the Li site in Li$_{0.9}$Mo$_{6}$O$_{17}$, as a function of ($\theta$, $\phi$) for the direction of $\vec{B_{0}}$ in space, plotted as $H/M$, and (b) $H_{\parallel}/M$, respectively. \label{fig5}}  
\end{figure}  

     In the following, the dipolar field in the Q1D paramagnetic conductor Li$_{0.9}$Mo$_{6}$O$_{17}$ is calculated.

     The crystal lattice\cite{onoda, john1} of Li$_{0.9}$Mo$_{6}$O$_{17}$ has a monoclinic space group $P2_{1}/m$, which has four equivalent sites in total for each site due to the symmetry of its 2-fold screw axis ($b$ is the default axis for the rotation). The space group also has an existence of a mirror plane, which is $\perp$ $b$. The four equivalent sites are \\
\\
($x'$, $y'$, $z'$) \\
($-x'$, $1/2 + y'$, $-z'$) \\
($-x'$, $-y'$, $-z'$) \\
($x'$, $1/2 - y'$, $z'$), \\
\\
i.e., each of which has the rest of three other equivalent sites to it. Here $x'$, $y'$, and $z'$ are the fractional coordinates in the lattice $abc$-system, and their relation with $x'_{a}$, $y'_{b}$, and $z'_{c}$ in Eq. (19) is, $x'_{a}$ = $ax'$, $y'_{b}$ = $by'$, and $z'_{c}$ = $cz'$, respectively.

     Figure 3 shows the crystal structure\cite{john1, john2} of Li$_{0.9}$Mo$_{6}$O$_{17}$. In each unit cell, there are six independent Mo sites, Mo1, Mo2, ..., Mo6, where the paramagnetic conduction electrons (with magnetic dipole moments) are from, while the number of independent sites for the Li (field observation site here) is only one. However, each site of them has three other sites, all of which are structurally equivalent to each other, as described above. Thus, in total there are 24 (6 $\times$ 4) Mo sites and 4 Li sites (again, all the Li sites are equivalent in the crystal structure) in each unit cell. Apparently, all the 24 Mo sites (each has a magnetic dipole moment) contribute to the dipolar fields at the Li sites, while the fields at the Li sites are the same. Even so, there will still be a large number of terms in the dipolar field calculations.   

     Table I shows the fractional coordinates $x'$, $y'$, and $z'$ for the positions of the independent Mo and Li sites in a unit cell (with the lattice constants) in the lattice coordinate system that are needed for the calculations. 

     With the matrix transform using Eq. (19), these fractional coordinates $x'$, $y'$, and $z'$ in a unit cell (for convenience, let's number it as $N'$ = 0) can be transformed into the Cartesian coordinates correspondingly. The same way needs to be applied to each neighboring unit cell ($N'$ = $\pm$1, $\pm$2, $\pm$3, ...) along $\pm ~\vec{e}_{x}$, $\pm ~\vec{e}_{y}$ and $\pm ~\vec{e}_{z}$, correspondingly, with the values of $x$, $y$, and $z$ to be used in Eqs. (8) - (10).  

     Figure 4 exhibits the calculated result of the magnetic dipole field components at the Li site due to the paramagnetic Mo electron moments in Li$_{0.9}$Mo$_{6}$O$_{17}$, plotted as $H_{x}/M$, $H_{y}/M$, and $H_{z}/M$ versus $\theta$ and $\phi$ (in arbitrary unit which leaves out the constant $\mu_{0}/4\pi$ in front) along the $x$, $y$, and $z$ directions, respectively, where $M$ is the magnitude of the magnetization, and $H_{x}$, $H_{y}$, and $H_{z}$ are the $x$, $y$, and $z$ components of the calculated magnetic dipole field intensity, respectively [Eqs. (6) - (10)]. 

     The calculation involves $N'$ = 10, i.e., $(2N'+1)$ = 21 unit cells along each axial ($x$, $y$ and $z$) direction (within a radius of $\sim$ 130 $\AA$). We also checked some calculations with $N'$ = 50 and 100, and the results are essentially the same. Noticeably, the values of $H_{x}/M$, $H_{y}/M$, and $H_{z}/M$ are completely determined by the lattice structure and the direction of $B_{0}$, while independent of the magnetic susceptibility $\chi$ and the magnitude of $B_{0}$.   

     As we can see from Fig. 4, each component including its minimum and maximum values has a rather strong angular dependence (i.e., the direction of $\vec{B}_{0}$). For example, for $B_{0}$ $\perp$ $b$ (i.e., $\phi$ = 0$^{\circ}$), the minimum for $H_{x}$, $H_{y}$, and $H_{z}$ is at $\theta$ = +340$^\circ$, +30$^\circ$, and +130$^\circ$ (angle $\theta_{\text{min}}$), respectively, while the corresponding maximum is at $\theta$ = +160$^\circ$, +210$^\circ$, and +310$^\circ$ (angle $\theta_{\text{max}}$), respectively. Thus, the angle difference between the maximum and minimum for each component always has $\theta$ = $|$$\pm$ 180$^\circ$$|$ (for $\phi$ = 0$^\circ$). Moreover, we also have the absolute values for the minimum and maximum for each component to be the same, i.e., $|$min($H_{k}$)$|$= $-$ $|$max($H_{k}$)$|$ (here $k$ = $x$, $y$, and $z$), and the minimum (maximum) for $H_{y}$ is the lowest among them (i.e., $H_{y}$ $<<$ $H_{x}$ and $H_{y}$ $<<$ $H_{z}$) when $\phi$ = 0$^\circ$. These are understandable because the paramagnetic electron moment tends to align along $B_{0}$, and $B_{0}$ is perpendicular to the $y$ ($b$) axis ($H_{y}$ component), i.e. $B_{0}$ is in the $xz$-plane which has the $H_{x}$ and $H_{z}$ components, when $\phi$ = 0$^\circ$. 
  
    On the other hand, that we have $|$min($H_{k}$)$|$ = $-$ $|$max($H_{k}$)$|$ here instead of $|$min($H_{k}$)$|$ = $-$ $|$max($H_{k}$)$|$/2 or $|$min($H_{k}$)$|$ = $-$ 2 $|$max($H_{k}$)$|$ indicates that there is no axial symmetry for the magnetic dipole field components from the Mo-electrons observing at the Li site in Li$_{0.9}$Mo$_{6}$O$_{17}$, as an axial symmetry would expect a $\pm$ (3cos$^{2}\theta$ $-$ 1) relation for a dipolar field as a function of $\theta$, which also requires the minimum of $H_{k}$ to be at + 54.7$^{\circ}$ ($\pm$ 180$^{\circ}$), an angle called the $``$magic angle'' ($\theta_{\text{magic}}$),\cite{abrag, jackson} i.e., $\theta_{\text{min}}$ = $\theta_{\text{magic}}$ $\equiv$ + 54.7$^{\circ}$ $\pm$ 180$^{\circ}$, corresponding to the angle $\theta$ that satisfies 3cos$^{2}\theta$ $-$ 1 = 0.   

    Figure 5 shows the result of the calculated magnetic dipole field component parallel to the externally applied magnetic field $B_{0}$ [Fig. 5(b)], as compared with the magnitude of the magnetic dipole field  [Fig. 5(a)] at the Li site, due to the paramagnetic Mo electron moments in Li$_{0.9}$Mo$_{6}$O$_{17}$. They are obtained based on the result shown in Fig. 4, plotted as $H_{\parallel}/M$ and $H/M$ versus $\theta$ and $\phi$, respectively, with mathematical expressions for them as, $H_{\parallel}$  = $\vec{H} \cdot \vec{B}_{0}$ = $H_{x} sin\theta cos\phi$ + $H_{y} sin\theta sin\phi$ + $H_{z} cos\phi$, and $H$ = $\sqrt{H_{x}^{2} + H_{y}^{2} + H_{z}^{2}}$.
\begin{figure}
\includegraphics[scale= 0.36]{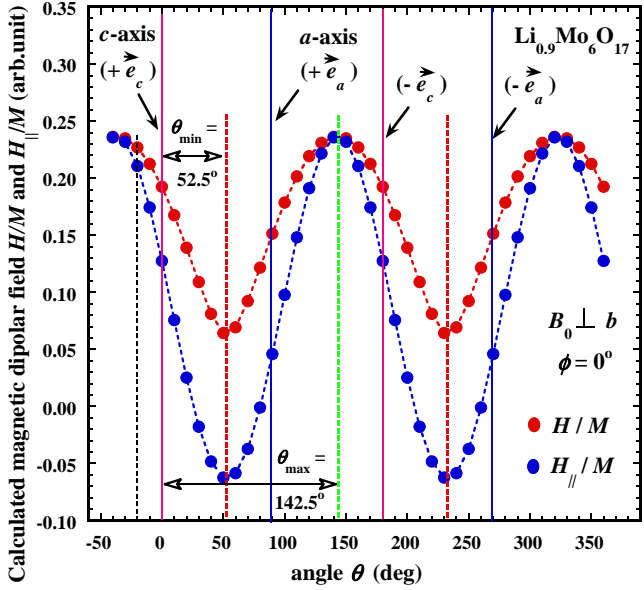}
\caption{(color online) Calculated magnetic dipole field magnitude ($H/M$) (red dots/curve) and magnetic dipole field component parallel to $\vec{B_{0}}$ ($H_{\parallel}$/M) (blue dots/curve) at the Li site in Li$_{0.9}$Mo$_{6}$O$_{17}$, as a function of angle $\theta$ for the direction of $\vec{B_{0}}$ when $\phi$ = 0 ($B_{0}$ $\perp$ $b$). The thin solid lines are labeled with arrows for the positions of the lattice $a$ and $c$ axes, and the thin dashed line (black color) at the angle $\theta$ = $-$ 20$^{\circ}$ indicates an angle that some of our experimental observations are at. The rest of the thick dashed lines indicate the angles at which the values of $H/M$ and $H_{\parallel}/M$ in minimum and maximum are. \label{fig6}}  
\end{figure}  

    Figure 5 indicates that $H_{\parallel}$ and $H$ have similar strong angular dependence with ($\theta$, $\phi$) to the axial dipolar field components $H_{x}$, $H_{y}$ and $H_{z}$, while their periods with both of the $\theta$ and $\phi$ dependence are essentially half of those for the axial dipolar field components, as we would expect since the values of the $H_{y}$ component here are generally a lot smaller than other components [Fig. (4)] as a major factor here. 

    One aspect for the importance of the $H_{\parallel}$ component is that $H_{\parallel}$ is the only component (i.e., $B_{||}^{\rm{dip}}$) in terms of the magnetic dipole field $\vec{H}$ that contributes to the Knight shift of a NMR spectrum, as mentioned earlier using $B_{||}^{\rm{dip}}$ ($\vec{B}$ $\parallel$ $\vec{B_{0}}$), where $B$ = $\mu_{0}H$ ($B_{||}^{\rm{dip}}$ = $\mu_{0}H_{\parallel}$), under the high field limit ($B$ $<<$ $B_{0}$).       

    Figure 6 shows the detailed values of $H_{\parallel}$ and $H$ for $\phi$ = 0 ($B$ $\perp$ $b$) as a function of $\theta$, which is exactly the case when the sample is set to rotate around the $b$-axis in the applied magnetic field $B_{0}$ as we had in our NMR experiments,\cite{wu1} while during the sample rotation $\vec{B}_{0}$ is kept in the $xz$-plane (also the $ac$-plane here for Li$_{0.9}$Mo$_{6}$O$_{17}$). 

    Interestingly, Fig. 6 indicates that 1) $H_{\parallel}$ has the same maximum value as $H$ [i.e., max($H_{\parallel}$/$M$) = max($H$/$M$)= $\sim$ +0.24 (arb.unit)], whereas their minimum values are very different, 2) the angles for their values in maximum are the same and the angles for their values in minimum are also the same. Their maximum and minimum values are at $\theta$ = $\sim$ $+$ 142.5$^{\circ}$ $\pm$ 180$^{\circ}$ ($\theta_{max}$) and $\theta$ = $\sim$ + 52.5$^{\circ}$ $\pm$ 180$^{\circ}$ ($\theta_{min}$), respectively, and 3) a range of angles for the small values of $H_{\parallel}$ are at the angles around the middle between the $a$ and $c$ axes in the $ac$-plane [i.e., at $\theta$ = $\sim$ (50 $\pm$ 30)$^{\circ}$] (the minimum is 7.5$^{\circ}$ closer to the $a$-axis than to the $c$-axis). In order words, at the angles closer to the $a$-axis (45$^{\circ}$ $<$ $\theta$ $\leq$ 90$^{\circ}$), the value of $H_{\parallel}$ is essentially negligible, and there is no axial symmetry (in terms of a dipolar field) for $H_{\parallel}$, either, observing from the Li site.   
\begin{figure}
\includegraphics[scale= 0.43]{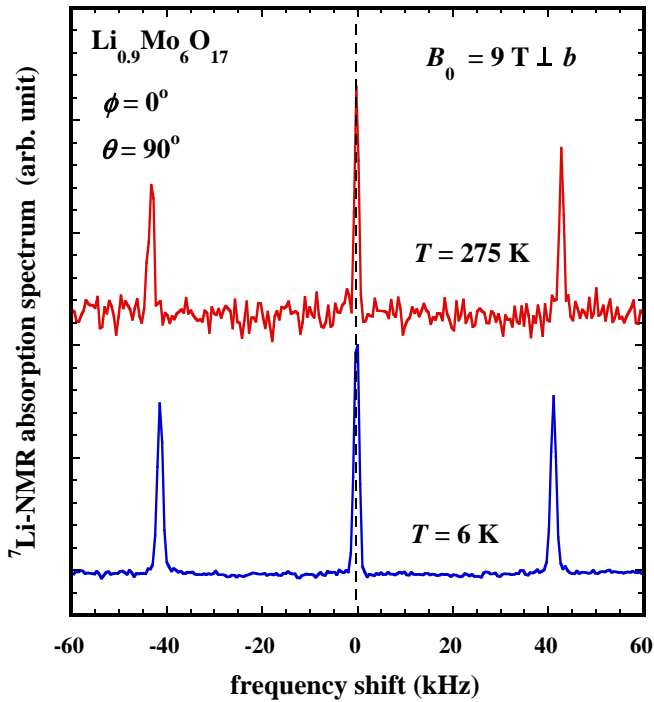}
\caption{(color online) Measured $^{7}$Li-NMR spectra of Li$_{0.9}$Mo$_{6}$O$_{17}$, with $B_{0}$ = 9 T $\perp$ $b$ (i.e., $\phi$ = 0$^{\circ}$) and $\sim$ along the lattice $a$-axis (i.e., $\theta$ = 90$^{\circ}$) at temperature $T$ = 275 K (upper red curve) and 6 K (lower blue curve). For comparison, the spectra are normalized and shifted on top of each other as indicated by the dashed line at the center (black color). Note, that the satellite peaks are lightly closer to the central line at 6 K than at 275 K is caused \cite{wu1} by a $\sim$ $\pm$ 1$^{\circ}$ angle change ($\Delta\theta$) due to the NMR sample probe thermal contraction upon cooling, with an extremely high sensitivity of the $^{7}$Li-NMR quadrupole frequency at $B_{0}$ $||$ $a$. \label{fig7}}  
\end{figure}  
\section{Result of the $^{7}$Li-NMR experimental observations}
%
%
    Figure 7 shows the result of the  measured $^{7}$Li-NMR spectra of Li$_{0.9}$Mo$_{6}$O$_{17}$, with $B_{0}$ = 9 T applied $\perp$ $b$ $\sim$ along the lattice $a$-axis, $\phi$ = 0$^{\circ}$ and $\theta$ = 90$^{\circ}$, at two typical temperatures $T$ = 275 K and 6 K, plotted as the spectrum absorption versus frequency shift (a shift from the NMR Larmor frequency $\nu_{0}$ = $^{7}\gamma_{I} B_{0}$ = 148.95 $\times$ 10$^{3}$ kHz here, where $^{7}\gamma_{I}$ = 16.547 MHz/T is the gyromagnetic ratio of the $^{7}$Li nucleus). The details for the measurements will ber published elsewhere \cite{wu1}.

    For comparison, the spectra are normalized to be 1 in a standard way for the intensity of the central line and shifted on top of each other. Since $^{7}$Li is a spin $I$ = 3/2 nucleus, theoretically, each $^{7}$Li-NMR spectrum is expected to have a central line plus two symmetric quadrupolar satellites due to the $^{7}$Li nucleus spin quantum $m$ = + 1/2 $\leftrightarrow$ $-$ 1/2 (central) and $\pm$ 3/2 $\leftrightarrow$ $\pm$ 1/2 (satellites) transitions, respectively. This is exactly what we experimentally see here, verifying that all the Li sites are structurally equivalent with a high quality sample being used \cite{wu1}. 

    Figure 7 shows that the width of the $^{7}$Li-NMR spectrum central line, more precisely, the full width half maximum (FWHM) of the central linewidth, has negligible change over a wide range of temperature from $T$ = 275 K to $T$ = 6 K, with a value of $\sim$ 0.6 kHz, including at the ``metal-insulator'' crossover temperature $T_{\text{MI}}$ = 24 K, i.e., there is no observable inhomogeneous magnetic broadening over a wide range of temperature (upon cooling from 275 K to 6 K), while during which range the sample magnetic susceptibility $\chi_{\text{DC}}(T)$ has a change of $\sim$ 2 times. This spectrum data indicates that the distribution of the local magnetic field parallel to $B_{0}$ has no change. 

    Since the field distribution is proportional\cite{abrag, slichter} to the sample dc magnetic susceptibility $\chi_{\text{DC}}(T)$ and the strength of the dipolar coupling between the $^{7}$Li nucleus and the Mo electrons $H_{||}/M$ ($B_{||}^{\text{dip}}$), which is the major source of the local magnetic field at the $^{7}$Li nucleus (see Sect. IV Discussion for details), this experimental spectrum data demonstrates that the value of $H_{||}/M$ ($B_{||}^{\text{dip}}$) is very small for $B_{0}$ $\parallel$ $a$ ($\phi$ = 0$^{\circ}$ and $\theta$ = 90$^{\circ}$) at the Li site, agreeing with the result of the theoretical calculations (Fig. 6).  

    This observation is further confirmed by our $^{7}$Li-NMR spectra versus angle data,\cite{wu1} for example, the correspondingly case at $\phi$ = 0$^{\circ}$ and $\theta$ = $-$ 20$^{\circ}$ or + 142.5$^{\circ}$, where the value of $B_{||}^{\text{dip}}$ is relatively large or has a maximum (near $B_{0}$ $\parallel$ $c$), is very different (not shown here). 

    On the other hand, a NMR spectrum line can have a mixture of many different sources of local electric and/or magnetic field contributions (Sect. IV D). According to the NMR theory \cite{slichter, abrag}, the satellites and the central line of the $^{7}$Li-NMR spectra have completely different origins: the central line is magnetic, due to the contribution of the nuclear spin interaction with the surrounding electron spins and other sources of the local magnetic fields, while the satellites are quadrupolar, coming from the contribution of the nuclear quadrupole moment interaction with the electric field gradient (EFG) due to the charges of the surrounding electrons (Mo electron charge contribution), i.e., the satellites are non-magnetic. Note, the quadrupolar interaction contribution (electric) to the central line is in the second order, thus having a negligible effect to the central line (non-electric). But its contribution to the satellites is in the first order, which is dominant (electric).

    However, a satellite can be both electronically and magnetically broadened. Thus, in terms of judging the charge contributions by an NMR spectrum satellite, it is important to separate or minimize any potential magnetic contributions in any way possible. That is exactly why we have the $^{7}$Li-NMR spectrum measurements at the direction $\sim$ along the $a$-axis (Fig. 7), where the magnetic dipole field has $\sim$ the smallest values (Fig. 6).

    Therefore, from the fact that Fig. 7 also shows that the satellites of the $^{7}$Li-NMR spectra including their FWHM width (the distribution of the EFG) have no changes, upon cooling in temperature, during which process the magnetic dipole field contributions to the satellites at $\theta$ = 90$^{\circ}$ is minimized (close to zero), we can see that there is no appearance of any charge effect anomaly during the cooling process, thus suggesting that the ``metal-insulator'' crossover at $T_{\text{MI}}$ = 24 K is not a charge effect.
\section{discussion}      
    In this section, we have discussions regarding the paramagnetic electron model used in the above calculations, the Pauli spin susceptibility ($\chi_{s}$) of the Mo electrons, the value of $B_{||}^{\rm{dip}}$ in Gauss (G) [at $B_{0}$ = 9 T , for example], the major sources of the local magnetic fields at the Li site, and possible effective magnetic dipole moment ($\mu_{\text{eff}}$) of the Mo electrons.    
\subsection{Paramagnetic electron model used in the calculation}
   In Sect. II, the magnetic dipole field in the quasi-one-dimensional (Q1D) paramagnetic conductor Li$_{0.9}$Mo$_{6}$O$_{17}$ is calculated when a single crystal sample is exposed to an externally applied magnetic field ($B_{0}$), using a paramagnetic electron model.

   The method involves a summation of the magnetic dipole fields from the average of the individual magnetic dipole moments of the electron spins in the crystal lattice in component forms, each of which converges with the increase of the number (up to 100) of the unit cells to be included, within a spherical distance centering at the observation atom (or nucleus) site. It also involves a general matrix transform corresponding to a fixed set-up of the Cartesian $x$, $y$, and $z$ axes relative to the $a$, $b$, and $c$ axes of the crystal lattice in the lattice coordinate system that can always be made. 

   Here we would like to point out that, in the paramagnetic electron model used for the above calculation, possible differences of the electron moments among the six independent Mo sites in the structure are not considered. Instead, we use their average magnetic dipole moments as reflected by the magnetization $M$ or magnetic susceptibility $\chi$ [Eqs. (5) - (6)]. We also neglect the individual electron interactions among the Mo electrons. Note, the interaction could polarize the electron dipole moments, which could also be reflected by the susceptibility data $\chi$ (see Sect. IV B).   

   Even so, it is still worthwhile to notice the possible difference in the electron moments and potential electron interactions, especially considering that some of the Mo electrons may not be equally conducting (or not conducting) according to their positions in the crystal lattice. In fact, along the $a$-axis in the crystal structure there are stacking layers (in the $bc$-plane) of Mo$_{4}$ octahedra separated by Mo$_{4}$ tetrahedra and the Li ions \cite{onoda, dumas}, and along the $b$-axis there is a double zig-zag Mo1-O-Mo4-O chain \cite{onoda, dumas, zspop}. Each chain involves only 2 (i.e., Mo1 and Mo4) out of 6 independent Mo sites (Mo1, Mo2,..., Mo6) that are believed \cite{zspop, xxu, john3} to have the electrons being the most conducting (conduction electrons), above the mysterious metal-insulator cross-over temperature at $T_{\text{MI}}$ = 24 K \cite{green, choi}. But below $T_{\text{MI}}$ there is a gradual dimensional crossover\cite{john1} and finally Li$_{0.9}$Mo$_{6}$O$_{17}$ becomes a 3D superconductor at $T$ $\leq$ 2.2 K upon cooling in temperature with $B_{0}$ = 0 or 0 $<$ $B_{0}$ $<$ $H_{c2}$ (upper critical field) \cite{lebed,jfm}.

   However, there is no clear evidence of anisotropy in the dc magnetic susceptibility ($\chi_{\text{DC}}$) which could reflect the difference in the electron moments and spin polarizations in Li$_{0.9}$Mo$_{6}$O$_{17}$, as the difference between the axial values of $\chi_{\text{DC}}$ that Matsuda $et ~al.$ showed \cite{matsuda} is actually rather small, which is very different from the high anisotropy character \cite{john3, jfm} in its electrical properties.
\subsection{Pauli spin susceptibility ($\chi_{s}$) of the Mo electrons}
    It is well-known that the Pauli spin (paramagnetic) susceptibility $\chi_{s}$ comes from the contributions of the conduction electron spin moments only,\cite{ashcroft} and the dc magnetic susceptibility ($\chi_{\text{DC}}$) has a general expression as $\chi_{\text{DC}}(T)$ = $\chi_{\text{dia}}$ + $\chi_{s}(T)$ + $\chi_{\text{orb}}$ + $\chi_{\text{other}}(T)$, where $\chi_{\text{dia}}$ and $\chi_{\text{orb}}$ are $T$-independent diamagnetic susceptibility and orbital susceptibility, respectively, and $\chi_{\text{other}}(T)$ comes from other sources, including the localized electron moments, lattice imperfection and/or impurities which could be also part of a Curie/Curie-Weiss paramagnetic contribution term and become dominant at low $T$.  

    The dc magnetic susceptibility ($\chi_{\text{DC}}$) measurements in Li$_{0.9}$Mo$_{6}$O$_{17}$ show that\cite{choi, musf}
\begin{eqnarray}
  \chi_{\text{DC}} = \frac{C}{T + \theta_{D}} + \chi_{01}, ~(T < 100 K),  \\
  \text{and} ~~ \chi_{\text{DC}} \approx (0.30 - 0.40)\times 10^{-4} ~~(\text{cm}^{3}/\text{mol.FU}), \nonumber \\
                       ~~~~~~~~~~~~ (100 K \leq T \leq 300 K), 
\end{eqnarray}
i.e., it has a Curie-Weiss susceptibility term appears at low temperatures ($T$ $<$ 100 K), where the Curie-Weiss constant $C$ = (7.8 $\pm$ 0.2) $\times $10$^{-4}$ cm$^{3}$$\cdot$K/mol.FU (note, FU $\equiv$ formula unit), $\theta_{D}$ = (6.1 $\pm$ 0,2) K, and $\chi_{01}$ = (0.181 $\pm$ 0.005) $\times$ 10$^{-4}$ cm$^{3}$/mol.FU. In the higt $T$ regime ($T$ $\geq$ 100 K), $\chi_{\text{DC}}$ has a value from $\sim$ 0.40 $\times$ 10$^{-4}$ cm$^{3}$/mol.FU at 300 K to $\sim$ 0.30 $\times$ 10$^{-4}$ cm$^{3}$/mol.FU at 100 K [Eq. (21)], i.e., it slowly decreases with a very weak $T$-dependence (close to linear here) upon cooling as expected for a quasi-1D conductor, where the $T$-dependence could have a contribution from the Pauli spin susceptibility $\chi_{s}(T)$ (note, $\chi_{s}$ has $T$-dependence for a 1D or 3D conductor) \cite{ach}.

    From the diamagnetism of the ions, we have $\chi_{\text{dia}}$ = $-$2.62 $\times$ 10$^{-4}$ cm$^{3}$/mol.FU, and according to the estimate \cite{matsuda} by Matsuda $et ~al.$, $\chi_{\text{orb}}$ $\approx$ 2.0 $\times$ 10$^{-4}$ cm$^{3}$/mol.FU. But it seems impractical to have further separations among the susceptibility data. 

    Most recent specific heat measurement\cite{jfm1} resulted in $\chi_{s}(T)$ at $T$ $\rightarrow$ 0 K as, $\chi_{s}(0)$ = 3.0 $\times$ 10$^{-6}$, i.e., 0.50 $\times$ 10$^{-4}$ cm$^{3}$/mol.FU (a factor with the molar density $\rho$ = 0.0595 mol/cm$^{3}$ for Li$_{0.9}$Mo$_{6}$O$_{17}$), using the measured Sommerfeld constant $\gamma_{S}$ = 1.6 mJ/mol.K$^{2}$ and the assumption of the Sommerfeld-Wilson ratio\cite{jfm1,merino} $R$ $\equiv$ 4$\pi^{2}k_{B}^{2}\chi_{s}(0)/[3(g\mu_B)^2\gamma_S]$ = 2, which applies for strongly correlated electrons and/or systems with repulsive interactions (here $\mu_{B}$ is the Bohr magneton, $k_{B}$ is the Boltzmann constant and $g$ is the Lande $g$-factor). Thus based on the measured values of $\chi_{\text{DC}}(T)$ [Eqs. (20)-(21)], we can also estimate the value of $\chi_{s}(T)$ at $T$ = 300 K, $\chi_{s}$(300 K) $\approx$ 3.6 $\times$ 10$^{-6}$, i.e. 0.60 $\times$ 10$^{-4}$ cm$^{3}$/mol.FU.   

    Noticeably, $\chi_{s}$ is just slightly larger than the high temperature value of $\chi_{\text{DC}}$. This is due to the cancellation of the orbital susceptibility $\chi_{\text{orb}}$ with the diamagnetic susceptibility $\chi_{\text{dia}}$, both of which are $\sim$ (5 $-$ 6) times larger than the high temperature value of $\chi_{\text{DC}}$. 

    Now, with the value of $\chi_{s}(0)$ we can find the density of state (DOS) $D(E_{F})$ at the Fermi energy ($E_F$) level \cite{ash, ach},
\begin{eqnarray}
   D(E_{F}) = \chi_{s}(0)/\mu_{B}^{2} \approx 1.5 ~~(\text{state/eV.FU}), \\
   \approx 0.25 ~~(\text{state/eV.ion}).
\nonumber
\end{eqnarray}
This value is close to the result obtained from the specific heat measurements, which is $D(E_F)$ = 3$\gamma_{S}Ne$/$(\pi^2k_{B}^2)$ $\approx$ 0.68 (\text{state/eV.ion}), where $N$ is the number of ions per unit cell, and $e$ is the electron charge \cite{ashcroft}.
 
   Correspondingly, the value of $E_{F}$ (at $T$ $\rightarrow$ 0) is \cite{ach} 
\begin{equation}
   E_{F} = d/[2D(E_{F})] \approx 6.0 ~~ (eV), \\
\end{equation}
where $d$ is the dimension for the conduction electrons. Here we had $d$ = 3 for Li$_{0.9}$Mo$_{6}$O$_{17}$ as it becomes a 3D conductor (superconductor) at $T$ $\rightarrow$ 0 K; both Eq. (22) and (23) are for 3D (not 1D) electrons. In comparison, this value of $E_F$ is slightly smaller than that of the free Cu electrons which has a value \cite{ashcroft} of $E_F$ = 7.8 eV.
\subsection{Value of $B_{||}^{\rm{dip}}$ in Gauss (G) at $B_{0}$ = 9 T} 
    Considering the unit (arbitrary) of $H_{||}/M$ shown in Figs. 5 - 6 and the Eqs. (6) - (10), we can have a convenient expression for $B_{||}^{\rm{dip}}$ as
\begin{equation}
    B_{||}^{\rm{dip}} = \frac{\mu_{0}}{4 \pi} \chi B_{0} \cdot (H_{||}/M) ~. \\
\end{equation}
Similar expressions can also be used for the dipolar field magnitude $B$ and the values of the corresponding axial components $B_{x}$, $B_{x}$ and $B_{x}$ ($\vec{B}$ = $\mu_{0}\vec{H}$), where the unit of $\vec{B}$ is in Gauss (G) or tesla (T) (1 T = 10$^{4}$ G).

     For example, at $B_{0}$ = 9 T with its direction angles $\phi$ = 0$^{\circ}$ and $\theta$ = 142.5$^{\circ}$, using the value of $H_{||}/M$ = 0.24 shown in Fig. 6 and the value of $\chi_{s}$(300 K) $\approx$ 0.6 $\times$ 10$^{-4}$ $\text{cm}^{3}/\text{mol.FU}$, equation (24) gives $B_{||}^{\rm{dip}}$ $\approx$ 0.35 G. This is the maximum (anisotropic) magnetic dipole field at the Li site that comes from the Mo-electron paramagnetic spins (magnetic dipole moments) in Li$_{0.9}$Mo$_{6}$O$_{17}$.  
\subsection{Major sources of the local magnetic fields at the Li site}
     The Hamiltonian ($H_{I}$) of the system for the $^{7}$Li-NMR in Li$_{0.9}$Mo$_{6}$O$_{17}$ can be expressed as \cite{slichter}    
\begin{equation}
  H_{I} = H_{IZ} + H_{II} + H_{Ie}^{\rm{Q}} + H_{Ie}^{\rm{dip}} + H_{Ie}^{\rm{contact}} + H^{\rm{demag}} + H^{\rm{Lor}}, \\ 
\end{equation}
where $H_{IZ}$ is the Zeeman Hamiltonian of the $^{7}$Li nucleus in $B_{0}$, $H_{II}$ is the $^{7}$Li-$^{7}$Li nuclear dipolar interaction Hamiltonian, $H_{Ie}^{\rm{Q}}$ is the Hamiltonian of the $^{7}$Li nuclear quadrupole interaction with the surrounding charges of the Mo electrons, $H_{Ie}^{\rm{dip}}$ and $H_{Ie}^{\rm{contact}}$ are the anisotropic dipolar hyperfine coupling and isotropic contact hyperfine to the Mo electron spins, respectively, and the last two terms, $H^{\rm{dem}}$ and $H^{\rm{Lor}}$, are the bulk demagnetization and Lorentz contributions, respectively \cite{slichter, carter}. Except for the first term $H_{IZ}$ which is for the Zeeman splitting of the $^{7}$Li nucleus's spin interaction with $B_{0}$, all of these terms contribute to the local magnetic or electric field at the Li site, contribute to the $^{7}$Li-NMR spectra and cause the NMR frequency shifts. 

    Noticeably, among these terms, only $H_{Ie}^{\rm{Q}}$ is non-magnetic \cite{slichter, abrag}; it has the first order (dominant) contribution to the local electric field (including the electric field distributions) at the $^{7}$Li sites, which can be fully reflected by the $^{7}$Li-NMR spectrum satellites. 

    Thus, by measuring the $^{7}$Li-NMR spectra and observing any potential changes of the spectrum satellites at the direction where the magnetic contributions (from the total of all the rest of the magnetic terms) to the satellites are minimized or zero, we can tell the behavior of the charges of the Mo conduction electrons precisely at the atomic scale, which is one of the major significances of this study.  

    Because the $^{7}$Li nucleus has a small atomic number $Z$ = 3, it is expected \cite{abrag} that its contact hyperfine couplings to the Mo electrons ($H_{Ie}^{\rm{contact}}$) is negligible. Thus, the system Hamiltonian can be re-written as 
\begin{equation}
   H_{I} \approx H_{IZ} + H_{II} + H_{Ie}^{\rm{Q}} + H_{Ie}^{\rm{dip}} + H^{\rm{dem}} + H^{\rm{Lor}}. \\
\end{equation}

    The term $H_{II}$ has a local magnetic field contribution ($B_{II}$) in the order of $^{7}\mu_{I}/r^{3}$, i.e., $B_{II}$ $\sim$ $^{7}\mu_{I}/r^{3}$, where $^{7}\mu_{I}$ is the spin moment of the $^{7}$Li nucleus, and $r$ is the distance between neighboring $^{7}$Li nuclei. Considering the value of $^{7}\mu_{I}$ = $^{7}\gamma_{I} \hbar I$ ($\hbar$ is the Planck's constant) and the minimum value of $r$ = 3.939 $\AA$ as well as the positions of $^{7}$Li in the crystal lattice, we have a rough estimate on $B_{II}$, which has an upper limit of $\sim$ 0.2 G. 

    Since $B_{II}$ is independent of temperature and unrelated to the Mo electron spins, it has no contribution to any potential line broadening of the $^{7}$Li-NMR spectra. Thus, our interest is in the last three terms, $H_{Ie}^{\rm{dip}}$, $H^{\rm{dem}}$, and $H^{\rm{Lor}}$, which are the terms related to the Mo electron spin dynamics and the local magnetic field properties at the Li site (again the term $H_{Ie}^{\rm{Q}}$ contributes to the local electric field only).

    In above Section IV C, we have estimated that at $B_{0}$ = 9 T along the lattice $a$-axis direction ($\theta$ = 90$^{\circ}$), $H_{Ie}^{\rm{dip}}$ has a dipolar field contribution $B_{||}^{\rm{dip}}$ $\approx$ 0.35 G at $T$ = 300 K. Now, we can estimate the magnetic field contributions of $H^{\rm{demag}}$ and $H^{\rm{Lor}}$ as \cite{carter}, $B^{\rm{demag}}$ = $-$ 4$\pi \cdot D \cdot \chi_{\text{DC}}(T)/(N_{\text{A}}\cdot \upsilon_{\text{Mo}}$), and $B^{\rm{Lor}}$ = + $4\pi /3 \cdot \chi_{\text{DC}}(T)/(N_{\text{A}}\cdot \upsilon_{\text{Mo}}$), respectively, where $D$ $\approx$ 0.45 is the estimated demagnetization factor along the $a$-axis according to the sample size, $N_{\text{A}}$ is the Avogadro's number, and $\upsilon_{\text{Mo}}$ = 669.5/24 $\AA^{3}$ is the unit cell volume per Mo ion (Table I). Thus, with the value of $\chi_{\text{DC}}(T)$ $\approx$ 0.4 $\times $10$^{-4}$ cm$^{3}$/mol.FU (at $T$ = 300 K) we have \cite{carter}
\begin{eqnarray}
   B^{\rm{demag}} + B^{\rm{Lor}} & = & 4\pi \cdot (\frac{1}{3} - D) \cdot \frac{\chi_{\text{DC}}(T)}{N_{\text{A}}\cdot \upsilon_{\text{Mo}}}, \\
                       & \approx & - ~0.05 ~(\text{G}).
\nonumber
\end{eqnarray}

    Therefore, the dipolar field of the Mo electron spins, i.e. the contribution of Hamiltonian $H_{Ie}^{\rm{dip}}$ (field $B_{||}^{\rm{dip}}$), is the dominant source of the local magnetic fields including the field dynamics at the Li site ($B_{||}^{\rm{dip}}$ $>$ $|B^{\rm{demag}}|$ and/or $B^{\rm{Lor}}$, and $B_{||}^{\rm{dip}}$ $>$ $B_{II}$). $B^{\rm{demag}}$ and $B^{\rm{Lor}}$ together here contribute little to the total local magnetic field at the Li site due to the very small value of $\chi_{\text{DC}}$ of the material, as evidence by the measured $^{7}$Li-NMR spectrum data (Fig. 7) (they have negligible impact on the spectra as the temperature varies).

    Note that any interaction among the Mo electrons in the crystal lattice will affect the polarization of the Mo electron spins, thereby modifying the dipolar field at the Li site from the Mo ions.
\subsection{Possible effective magnetic dipole moment ($\mu_{\text{eff}}$) of the Mo electrons}
    There are important studies\cite{clog} regarding how to obtain the possible effective magnetic dipole moment ($\mu_{\text{eff}}$) of the conduction electrons as described in Ref. [45]. Similar method was used recently in Ref. [46], which supposes that the value of $\mu_{\text{eff}}$ is proportional to the spin susceptibility $\chi_{s}$ (both $\mu_{\text{eff}}$ and $\chi_{s}$ can be $T$-dependent) \cite{grin},
\begin{equation}
   \mu_{\text{eff}}(T) = \frac{\mu_{\text{eff}}(T_{\text{ref}})}{\chi_{s}^{\text{ref}}(T_{\text{ref}})} \chi_{s}(T),  ~~ (T > T_{\text{ref}}) \\
\end{equation}
where $T_{\text{ref}}$ is the reference temperature low or high enough to obtain the local moment (a matrix element with a $T$-dependent susceptibility\cite{clog}), and $\chi_{s}^{\text{ref}}(T_{\text{ref}})$ and $\mu_{\text{eff}}(T_{\text{ref}})$ are the spin susceptibility and local moment (effective moment) at $T_{\text{ref}}$, respectively.  

    For Li$_{0.9}$Mo$_{6}$O$_{17}$ we choose $T_{\text{ref}}$ = 100 K, the highest temperature at which the Curie-Weiss type paramagnetic susceptibility applies here [Eq. (20)] [note, above 100 K it is the Eq. (21) that describes the susceptibility]. Since $\chi_{\text{eff}}(T_{\text{ref}}$ = 100 K) $\approx$ 0.53 $\times$ 10$^{-4}$ $\text{cm}^{3}/\text{mol.FU}$, $\chi_{\text{eff}}(T = 300 K)$ $\approx$ 0.60 $\times$ 10$^{-4}$ $\text{cm}^{3}/\text{mol.FU}$, and from the Curie-Weiss constant $C$ $\approx$ 7.8 $\times$ 10$^{-4}$ cm$^{3}$$\cdot$K/mol.FU (Section IV B) we have $\mu_{\text{eff}}$($T_{\text{ref}}$ = 100 K) = 2.82 $\sqrt{C}$ $\approx$ 0.08 $\mu_{B}$/FU = 0.08 $\mu_{B}$/(6 Mo ions) $\approx$ 0.013 $\mu_{B}$ per Mo-ion, equation (28) gives $\mu_{\text{eff}}$($T$ = 300 K) $\approx$ 0.015 $\mu_{B}$ per Mo-ion, as a possible effective magnetic dipole moment of the Mo electrons. 

    This value $\mu_{\text{eff}}$ is $\sim$ 100 times smaller than that of a spin $S$ = 1/2 free electron, which has a magnetic dipole moment $\mu_{\text{free}}$ = $\sqrt{4S(S+1)}$ $\mu_{B}$ = 1.73 $\mu_{B}$. The cause for this very small value of $\mu_{\text{eff}}$ for the Mo electrons in Li$_{0.9}$Mo$_{6}$O$_{17}$ is not clear \cite{choi}. On the other hand, any magnetic or non-magnetic impurities could also contribute to a local moment (especially at low temperatures) \cite{grin, alloul}, but there has been no evidence to show that the low temperature Curie-Weiss susceptibility term [Eq. (20)] is from any magnetic or non-magnetic impurities, while there is no lattice imperfection here.   
\section{Conclusions}
   The magnetic dipole field at the atomic scale in a single crystal of quasi-one-dimensional (Q1D) paramagnetic conductor Li$_{0.9}$Mo$_{6}$O$_{17}$ is investigated both theoretically and experimentally, using a paramagnetic electron model and $^{7}$Li-NMR spectroscopy measurements with an externally applied magnetic field $B_{0}$ = 9 T. The method is described in a general form and the field in the Li$_{0.9}$Mo$_{6}$O$_{17}$ is calculated as a function of the orientation angles ($\theta$ and $\phi$) of $B_{0}$ in space, with experimental observations $\&$ demonstrations. 

   We find that the magnetic dipole field component $B_{||}^{\text{dip}}$ parallel to $B_{0}$ has no lattice axial symmetry; it is the smallest around the middle between the lattice $a$ and $c$ axes, with the central minimum to be 7.5$^{\circ}$ closer to the $a$ than to the $c$ axis in the $ac$-plane, while the maximum of $B_{||}^{\text{dip}}$ is 51.9$^{\circ}$ from the $a$-axis on the other side (in the same $ac$-plane), with a maximum value $\sim$ 0.35 G when $B_{0}$ = 9 T $\perp$ $b$. The Mo ions could have a very small effective magnetic dipole moment $\mu_{\text{eff}}$ of 0.015 $\mu_{\text{B}}$. 

   By minimizing potential magnetic contributions to the $^{7}$Li-NMR spectrum satellites with the NMR spectroscopy measurements at the direction where the value of the magnetic dipole field is $\sim$ the smallest ($\sim$ $B_{0}$ $\parallel$ $a$; $\theta$ = 90$^{\circ}$ and $\phi$ = 0$^{\circ}$), the behavior of the independent charge contributions is observed: there is no $^{7}$Li-NMR spectrum satellite line broadening, i.e, no change in the charge distribution (including the value of EFG) from the Mo conduction electrons upon cooling over a wide range of temperatures.

  Other related important physics quantities such as the spin susceptibility $\chi_{s}$, DOS $D(E_{F})$, and the Fermi energy $E_{F}$ of the Mo electrons are also discussed.

   This investigation demonstrates that the magnetic dipole field from the Mo electrons is the dominant source of the local magnetic fields at the Li site, and by the measurements of the $^{7}$Li-NMR spectra at the direction where the value of the magnetic dipole field is $\sim$ the smallest, we are able to observe the behavior of the charge contributions of the Mo conduction electrons directly at the atomic scale. This investigation suggests that the mysterious ``metal-insulator'' crossover at low temperatures is fundamentally not a charge effect. The work also reveals valuable local field information for further NMR investigation as recently suggested to be key important to the understanding of many mysterious properties of this Q1D material.   
\begin{acknowledgments}
     The work at University of West Florida was supported by SCA-2012 (G. Wu) and at UCLA by NSF Grant DMR-0334869 (WGC). We thank J. L. Musfeldt, J. R. Thompson, J. J. Neumeier, and S. E. Brown for helpful discussions.  
\end{acknowledgments}

\end{document}